\documentclass[aip,rsi,reprint,graphicx]{revtex4-1} 

\usepackage{color}
\usepackage[dvips]{graphicx}
\usepackage{amsmath}
\usepackage{bm}

\draft 

\begin{document}


\title{Adiabatic measurements of magneto-caloric effects in pulsed high magnetic fields up to 55 T} 



\author{T. Kihara, Y. Kohama, Y. Hashimoto, S. Katsumoto, and M. Tokunaga}
\affiliation{The Institute for Solid State Physics, The University of Tokyo, Kashiwa 277-8581, Japan}


\date{\today}

\begin{abstract}
Magneto-caloric effects (MCEs) measurement system in adiabatic condition is proposed to investigate the thermodynamic properties in pulsed magnetic fields up to 55 T. With taking the advantage of the fast field-sweep rate in pulsed field, adiabatic measurements of MCEs were carried out at various temperatures. To obtain the prompt response of the thermometer in the pulsed field, a thin film thermometer is grown directly on the sample surfaces. The validity of the present setup was demonstrated in the wide temperature range through the measurements on Gd at about room temperature and on Gd$_3$Ga$_5$O$_{12}$ at low temperatures. The both results show reasonable agreement with the data reported earlier. By comparing the MCE data with the specific heat data, we could estimate the entropy as functions of magnetic field and temperature. The results demonstrate the possibility that our approach can trace the change in transition temperature caused by the external field.
\end{abstract}


\maketitle 

\section{Introduction}
The magneto-caloric effect (MCE) is a temperature change of magnetic material through the application or removal of an external magnetic field. This effect is a consequence of the field variation of the entropy and has been of considerable interest for their potential applications and underlying physics~\cite{rad,jai1,jai2,sil,acz,acz2,koh,ros,for}. The measurement of MCE has long been used to map out the magnetic phase diagram, and the combination with the specific heat measurement allows us to determine the order of the phase transition~\cite{sil} and the field evolution of entropy~\cite{ros,acz,acz2}. Because of the unique abilities of the MCE measurements, a lot of experimental techniques have been developed~\cite{jai3,sil,acz,acz2,koh,lev,dan1}.

In spite of the potential to extend the accessible magnetic field range, the MCE measurements in a pulsed field have been limited by the difficulties of the thermometry in the short pulse duration. One of the possible way to obtain MCE data in pulsed field was described by Levitin \textit{et al}~\cite{lev}. Here they measured the adiabatic magnetization curve of Gd$_3$Ga$_5$O$_{12}$ (GGG) in the extremely short pulse duration of 9 ms and compared it with the calculated isothermal magnetization curve. Since the difference between the adiabatic and the isothermal magnetization curves stems from the change in temperature as a function of magnetic fields, the comparison between them yields the MCE indirectly, as long as the calculation of the isothermal magnetization curve is valid. On the other hand, by applying a traditional way of the thermometry used thermocouple, Dan'kov \textit{et al.} directly measured the adiabatic MCE in pulse magnetic fields up to 8T and in the temperature range between 200 K and 350 K~\cite{dan1}. Here, they embedded a thin copper-constantan thermocouple in a polycrystalline specimen and achieved the rapid response of the thermometry necessary for a pulsed filed experiment. Although a thermocouple is an adequate choice for the thermometry in a pulsed field, the accessible ranges of temperatures and magnetic fields are restricted by the sensitivity of the thermocouple. More recently, the direct MCE measurement in the quasi-adiabatic limit was carried out below 8 K and magnetic field up to 60 T~\cite{koh,acz2}. This technique used a small chip of a resistive thermometer and achieved sufficiently fast response time for the thermometry while maintaining its sensitivity. However, because of the heat loss during the measurement, the quasi-adiabatic MCE experiment in the pulsed field is unable to obtain the field variation of the entropy.

In this context, we design a new apparatus for an accurate direct measurement of the adiabatic MCE over a wide range of temperatures and magnetic fields. The temperature is measured by a resistive film thermometer fabricated on the sample surface, and the apparatus was tested through the measurements of Gd and GGG. The results agree with the previous reports, and the experimental technique described in this article can be used to measure the adiabatic MCE in pulse magnetic fields up to 55 T in the temperature region between 6 K and 300 K.

\section{Principle underlying adiabatic MCE measurement}
Let us start with the following thermodynamic equation for a reversible process,
\begin{equation}
\delta q=TdS=T\left(\frac{\partial S}{\partial T}\right)_HdT+T\left(\frac{\partial S}{\partial H}\right)_TdH. \label{q}
\end{equation}
Here, $\delta q$ represents the heat exchanged with the surroundings. $S$, $T$, and $H$ denote the entropy, temperature and magnetic field, respectively. The first term of the right-hand side can be expressed as $C_HdT$ by using the heat capacity at the fixed field ($C_H$). On the other hand, the second term corresponds to the released heat caused by the field-induced change in the entropy. In the real system, we have to take into account finite heat exchange between the sample and the thermal bath. With introducing the thermal conductance $\kappa_{\textrm{b}}$, the heat balance equation is as follows:
\begin{equation}
\kappa_{\textrm{b}}(T_{\textrm{s}}-T_{\textrm{b}})=C_{\textrm{s}}\frac{dT_{\textrm{s}}}{dt}+T_{\textrm{s}}\left(\frac{\partial S}{\partial H}\right)_{T_{\textrm{s}}}\frac{dH}{dt}. \label{Cs}\\
\end{equation}
Here, the subscripts of "s" and "b" represent the sample and the thermal bath, respectively. If the $dH/dt$, and the $dT_{\textrm{s}}/dt=dT_{\textrm{s}}/dH\times dH/dt$ as well, is large enough, the heat leak term in the left-hand side of the Eq. \ref{Cs} became negligible: the adiabatic condition can be effectively accomplished. Hence, the fast field-sweep rate of the pulsed field is advantageous to realize the adiabatic measurements. In the purely adiabatic condition [$\delta q=0$ in Eq. \ref{q} or $\kappa_{\textrm{b}}(T_{\textrm{s}}-T_{\textrm{b}})=0$ in Eq. \ref{Cs}], $dT$ in Eq. \ref{q} is expressed as $(\partial T/\partial S)_H(\partial S/\partial H)_TdH=-(\partial T/\partial H)_SdH$. Therefore, the MCE [$\Delta T_{\textrm{ad}}(H_1\rightarrow H_2)$] can be expressed as
\begin{align}
\Delta T_{\textrm{ad}}(H_1\rightarrow H_2) &= \int_{H_1}^{H_2} \left(\frac{\partial T}{\partial H}\right)_{S}dH \notag \\
&= -\int_{H_1}^{H_2} \frac{T}{C_{\textrm{s}}}\left(\frac{\partial S}{\partial H}\right)_{T_{\textrm{s}}}dH. \label{dT}
\end{align}

This equation expresses the fact that the system temperature changes isentropically when the measurement is performed in an adiabatic condition. Therefore, the adiabatic MCE measurement monitors temperature at which the total entropy of the system remains constant as a function of magnetic field. For example, when the spin entropy of the system decrease by the application of the external magnetic field as seen in the paramagnetic system (negative $\partial S/\partial H$ in Eq. \ref{dT}), the system attempts to keep the total entropy, and the temperature of the system is raised. The opposite situation (positive $\partial S/\partial H$ and negative MCE) can occur in the typical antiferromagnetic system~\cite{kum}, where the antiferromagnetically ordered (low entropy) state is destroyed by the application of the magnetic field, leading to the negative MCE. In this manner, the sign of the MCE relates to the underlying magnetic structure.

\begin{figure}
\centering
\includegraphics[width=8.5cm,clip]{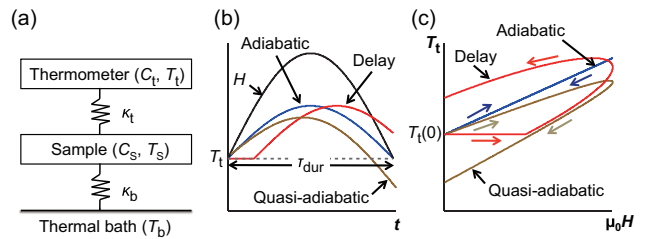}
\caption{(Color online) (a) Model for representing the MCE measurement set-up. The sample is coupled to the bath and the thermometer by the thermal conductances $\kappa_{\textrm{b}}$ and $\kappa_{\textrm{t}}$, respectively. Schematics of the changes in $H$ and $T_{\textrm{t}}$ as functions of (b) $t$ and (c) $\mu_0H$.}
\label{ad}
\end{figure}

Figure \ref{ad} (a) shows the schematic drawing of an actual MCE set-up. Here, $C_{\textrm{t}}$ and $T_{\textrm{t}}$ are the heat capacity and temperature of the thermometer. The thermometer strongly connects to the sample, and its thermal conductance is defined as $\kappa_{\textrm{t}}$. The sample is mounted on the thermal bath via a weak thermal link with the thermal conductance, $\kappa_{\textrm{b}}$. The link leads to heat exchange between the sample and the thermal bath which is expressed by the left-hand side of Eq. \ref{Cs}. This configuration of Fig. \ref{ad} (a) has been studied for modeling the relaxation and quasi-adiabatic calorimeters~\cite{wil,las}. When the conditions of $C_{\textrm{s}}\gg C_{\textrm{t}}$ and $\kappa_{\textrm{b}}\ll \kappa_{\textrm{t}}$ are satisfied, it is known that the time constant to reach temperature homogeneity between the sample and thermal bath is given by $\tau_{\textrm{b}}=C_{\textrm{s}}/\kappa_{\textrm{b}}$. In the same conditions, the thermal time constant between the sample and thermometer is given by $\tau_{\textrm{t}}=C_{\textrm{t}}/\kappa_{\textrm{t}}$. When the time scale of the MCE measurement is much shorter than $\tau_{\textrm{b}}$, the system cannot exchange the heat to the surroundings which is regarded as an adiabatic condition. The time scale of the MCE measurement can be roughly approximated by the pulsed field duration, $\tau_{\textrm{dur}}$, and the requirement condition for reaching an adiabatic condition during the measurement is given by $\tau_{\textrm{b}}\gg \tau_{\textrm{dur}}$. When the $\tau_{\textrm{t}}$ is longer than the characteristic time of the measurement ($\tau_{\textrm{dur}}$), the thermometer cannot respond to the change in $T_{\textrm{s}}$. Thus, to measure $T_{\textrm{s}}$ in the pulsed magnetic field, It is necessary to satisfy $\tau_{\textrm{dur}}\gg \tau_{\textrm{t}}$, and as the result, the following condition, $\tau_{\textrm{b}}\gg \tau_{\textrm{dur}}\gg \tau_{\textrm{t}}$, is required for the adiabatic MCE measurement. When the criterion of $\tau_{\textrm{dur}}\gg \tau_{\textrm{t}}$ is violated ($\tau_{\textrm{b}}\gg \tau_{\textrm{dur}}\sim \tau_{\textrm{t}}$), one can be expected to observe the "Delay" response as shown in Fig.\ref{ad} (b). When the $\tau_{\textrm{b}}$ is compatible with $\tau_{\textrm{dur}}$ ($\tau_{\textrm{b}}\sim \tau_{\textrm{dur}}\gg \tau_{\textrm{t}}$), the system is recognized as a "Quasi-adiabatic" condition. Both "Delay" and "Quasi-adiabatic" curves cannot be symmetric between the field-up and field down sweeps, and the hysteretic $T_{\textrm{t}}-H$ curves are observed as seen in Fig. \ref{ad} (c). On the contrary, the adiabatic MCE data are known to show no hysteretic $T_{\textrm{t}}-H$ curves~\cite{koh}, and the reliability of the adiabatic MCE measurement can be checked by the observation of the close loop in the $T_{\textrm{t}}-H$ curve.

It should be noted that we apply a thin film thermometer to obtain a rapid response of the thermometry ($\tau_{\textrm{dur}}\gg \tau_{\textrm{t}}=C_{\textrm{t}}/\kappa_{\textrm{t}}$), while the use of the pulsed magnetic field helps to achieve the adiabatic condition of $\tau_{\textrm{b}}=C_{\textrm{s}}/\kappa_{\textrm{b}}\gg \tau_{\textrm{dur}}$. To reach the condition of $\tau_{\textrm{dur}}\gg \tau_{\textrm{t}}$, the earlier works choose a thermometer having a small $C_{\textrm{t}}$ such as a thermocouple and a chip thermometer~\cite{dan1,koh}. In this point of view, the $C_{\textrm{t}}$ of the thin film thermometer is negligible small due to the thickness of a film thermometer which sufficiently reduces the $\tau_{\textrm{t}}$ for the pulsed field experiments.

\section{Experimental setup}
\begin{figure}[tbp]
\centering
\includegraphics[width=8.5cm,clip]{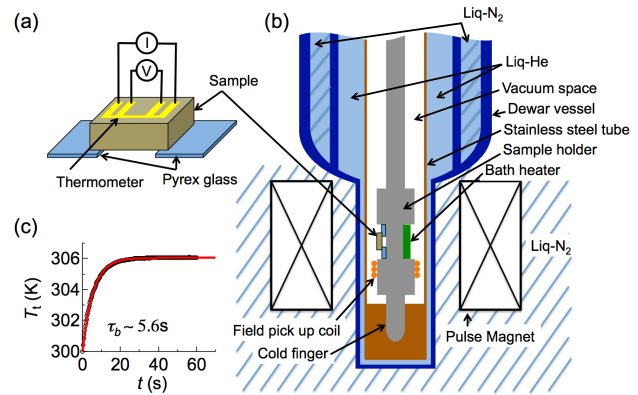}
\caption{(Color online) Schematic drawing for (a) the sample with thermometer and (b) the probe setup. (c) Relaxation curve at 300 K measured by applying a heating current to the film thermometer as a heater.}
\label{pr}
\end{figure}

Figures \ref{pr} show schematic drawings of (a) the sample with the thermometer and (b) the probe setup. 
For the MCE measurements at high temperatures, we chose Gd as a test sample because magnetic and thermal properties of Gd have been studied in detail at about room temperature~\cite{dan1,dan2}. The polycrystalline Gd sample with a purity of 99.5 wt.\% was produced by Wako Chemicals. In order to reduce the eddy current heating induced by the application of the pulsed field, the samples were shaped into the thin plates with the dimension of about 2 mm $\times$ 2 mm $\times$ 0.1 mm. The magnetic fields were applied parallel to the sample plane. To electrically isolate the film thermometer with the metallic Gd sample, the 150 nm-thick CaF$_2$ insulating layer was formed by the vacuum evaporation technique. Then, a patterned Au thermometer was deposited as shown in Fig. \ref{pr}(a) through a metallic mask. The dimensions of the sensing area were 1 mm $\times$ 0.1 mm $\times$ 100 nm. The heat capacity of this film thermometer $C_{\textrm{t}}$ is estimated as 0.5 $\mu$J/K at 300 K, which is negligibly small in comparison with that of the Gd sample ($C_{\textrm{s}}\sim500$ $\mu$J/K at 300 K). To measure the resistance of the thermometer, four Au wires with diameter of 0.05 mm were connected to the Au film thermometer using Ag-paste at where the contact is more than 0.5 mm far from the temperature sensing area.

As a test of the measurement system at low temperatures, we performed the adiabatic 
MCE measurement on the GGG single crystalline sample produced by Furuuchi Chemical Corporation. The dimension of the sample is 7 mm $\times$ 5 mm $\times$ 0.5 mm. The heat capacity of the sample was estimated to $C_{\textrm{s}}\sim100$ $\mu$J/K at 4.2 K. This value is sufficiently larger than $C_{\textrm{t}}\sim$ 0.13 nJ/K, which is calculated from the specific heat of Au$_\textrm{0.18}$Ge$_\textrm{0.82}$ at 4.2 K~\cite{moll}. The magnetic fields were applied parallel to the (111) plane. We used an Au$_{x}$Ge$_{1-x}$ ($x=0.18$ at.\%) alloy film grown on the sample surface by the radio-frequency sputtering. Since the temperature dependences of the resistivity in Au$_{x}$Ge$_{1-x}$ alloys depend strongly on the composition ratio $x$~\cite{dod,sah}, we can tune the sensitivity of the film thermometer depending on the measurement temperature range. The dimensions of the sensor in the Au$_\textrm{0.18}$Ge$_\textrm{0.82}$ film thermometer is 0.1 mm $\times$ 0.5 mm $\times$ 100 nm.
In order to reduce the $\kappa_{\textrm{b}}$, the sample with the film thermometer was fixed on the low thermal conducting Pyrex$^{\text{\textregistered}}$ glass plates by epoxy resin (Stycast$^{\text{\textregistered}}$ 1266). The probe (assembly of the sample, the film thermometer, the sample holder, bath heater, etc) was inserted in the thin-walled tube made of non-magnetic stainless steel as shown in Fig. \ref{pr}(b). The inside of the tube was evacuated to reduce the heat exchange between the sample and the thermal bath by the residual gas. 
 The resistance of the film thermometer was measured by the standard dc four-probe method or ac method using numerical lock-in technique at a frequency of 50 kHz\cite{koh}.

To evaluate the $\tau_{\textrm{b}}$, we applied a large amount of current to the Au film thermometer on the sample as a heater and measured the temporal variation of sample temperature at 300 K. As shown in Fig. \ref{pr}(c), the relaxation curve of $T_{\textrm{t}}(t)$ can be fitted well by single exponential function. In the MCE measurement of Gd, the $\tau_{\textrm{b}}$ was estimated as 5.6 s. This value is much longer than the total duration of our pulsed field (36 ms). Therefore, our system can be regarded to be in the adiabatic condition during the pulsed field application. We also estimated the $\tau_{\textrm{b}}$ for the GGG sample at 4.2 K and found that the $\tau_{\textrm{b}}$ of 500 ms is more than ten times greater than the $\tau_{\textrm{dur}}$. It indicates that our system is available for the adiabatic MCE measurements also at low temperatures.


\section{Results and Discussion}
\begin{figure}[tbp]
\centering
\includegraphics[width=8cm,clip]{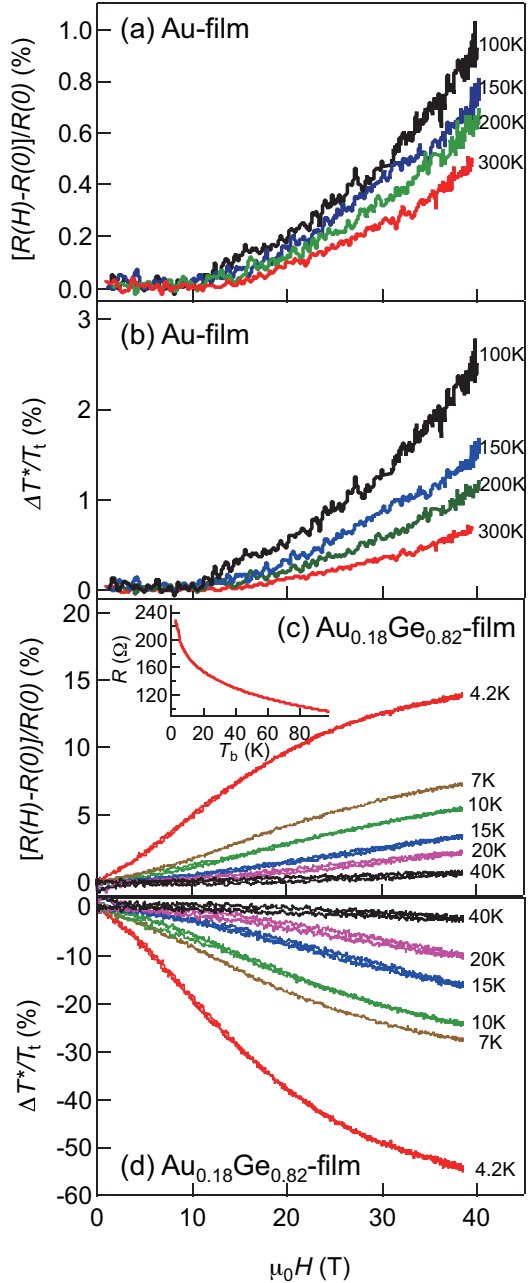}
\caption{(Color online) Magnetoresistance of (a) the Au film thermometer deposited on a sapphire substrate and (c) the Au$_{0.18}$Ge$_{0.82}$ film thermometer deposited on a quartz substrate, Inset: Temperature dependence of R without magnetic field. Artificial temperature change $\Delta T_{\textrm{t}}^*$ caused by the magnetoresistance of (b) the Au film thermometer and (d) the Au$_{0.18}$Ge$_{0.82}$ film thermometer.}
\label{au}
\end{figure}

First, to calibrate the film thermometers grown on the samples at zero field, we measured the zero-field resistance ($R$) as a function of $T_{\textrm{b}}$. In these experiments, we introduced sufficient amount of He gas to the sample space to achieve the thermal equilibrium between the sample and the thermal bath. The resistance of the metallic Au film shows slightly sub-linear temperature dependence with the slopes of $(dR/dT)/R=3.7\times10^{-3}$ K$^{-1}$ and 2.1$\times10^{-3}$ K$^{-1}$ at 100 K and 300 K, respectively. To compensate the effects of magnetoresistance in the film thermometers, we measured longitudinal magnetoresistances (MR) of these films grown on sapphire substrates with the similar process. Since non-magnetic sapphire does not show the MCE, we can evaluate the contribution of the isothermal magnetoresistance effects. The results are shown in Fig. \ref{au}(a). If we convert the resistance to temperature using the $R-T$ curve at zero-field, the MR effect at 40 T causes artificial temperature change of 2.4 K and 2.0 K at 100 K and 300 K, respectively [Fig. \ref{au}(b)]. On the other hand, the Au$_{0.18}$Ge$_{0.82}$ film shows semiconducting $R-T$ curve as shown in the inset of Fig. \ref{au}(c), and has higher sensitivity as a thermometer of $(dR/dT)/R=-5.9\times10^{-2}$ K$^{-1}$ and $-6.9\times10^{-3}$ K$^{-1}$ at 4.2 K and 40 K, respectively. The longitudinal MR of the Au$_{0.18}$Ge$_{0.82}$ film shows somewhat complicated positive behavior [Fig. \ref{au}(c)]. This leads to artificial change in temperature of $-2.3$ K at 4.2 K and $-0.9$ K at 40 K in the field of 38 T.

\begin{figure}[tbp]
\centering
\includegraphics[width=7.9cm,clip]{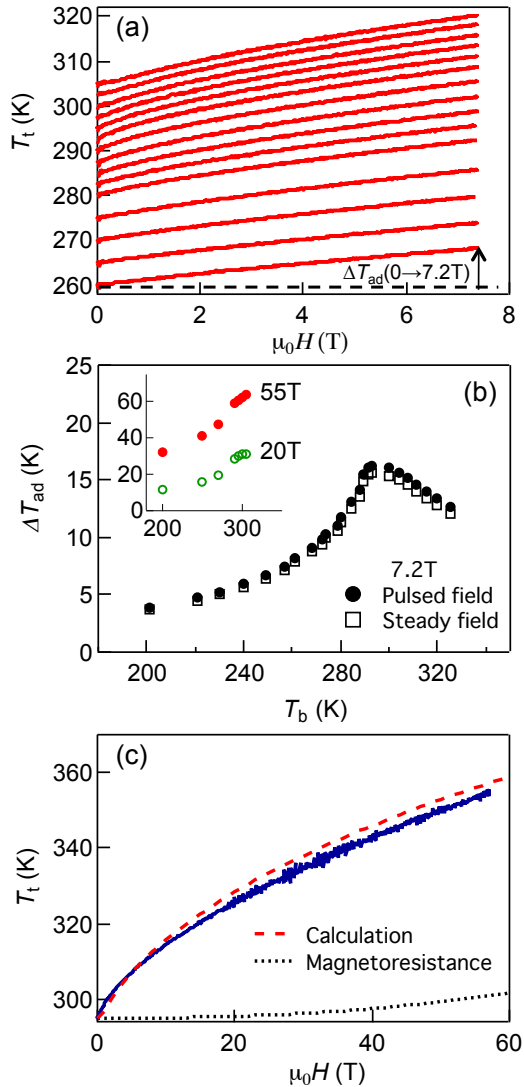}
\caption{(Color online) (a) $H$ dependence of $T_{\textrm{t}}$ on the Gd sample measured in pulsed magnetic fields up to 7.2 T at various $T_{\textrm{b}}$. The arrow corresponds to $\Delta T_{\textrm{ad}}(0\rightarrow7.2\textrm{T})$. (b) $\Delta T_{\textrm{ad}}(0\rightarrow7.2\textrm{T})$ vs $T_{\textrm{b}}$. Solid circles: $\Delta T_{\textrm{ad}}(0\rightarrow7.2\textrm{T})$ measured in the pulsed fields, Open squares: the data from Ref. 14 which were estimated by the specific heats measured in steady field. The data from Ref. 14 are interpolated from 7.5 T to 7.2 T. Inset: $\Delta T_{\textrm{ad}}(0\rightarrow20\textrm{T})$ and $\Delta T_{\textrm{ad}}(0\rightarrow55\textrm{T})$ as a function of $T_{\textrm{b}}$. (c) $T_{\textrm{t}}$ vs $H$ up to 55 T at $T_{\textrm{b}}$ = 295 K (blue line), Red dashed curve: the calculated data by molecular field approximation and Eq. (\ref{dT}), Black dotted curve: the artificial temperature change caused by the MR in Au film thermometer.}
\label{gd}
\end{figure}

To evaluate the validity of our measurement system, we measured the MCEs of Gd at about room temperatures. The Gd is a conventional ferromagnet with a large angular momentum of $J=7/2$ and Curie temperature $T_{\textrm{C}}$ of 293 K\cite{dan1,dan2}. Figure \ref{gd}(a) shows the results of $H$ dependence of the $T_{\textrm{t}}$ measured in pulsed magnetic fields up to 7.2 T. The $T_{\textrm{t}}-H$ curves in heating (field-increasing) and cooling (field-decreasing) processes coincide with each other. These reversible data in Fig. \ref{gd}(a) confilm that the condition of $\tau_{\textrm{b}}\gg \tau_{\textrm{dur}}\gg \tau_{\textrm{t}}$ is successfully satisfied in the pulsed field experiments. Therefore, we regard as $T_{\textrm{t}}=T_{\textrm{s}}$ in the following.

Figure \ref{gd}(b) exhibits the temperature dependence of $\Delta T_{\textrm{ad}}(0\rightarrow H)$. The $\Delta T_{\textrm{ad}}(0\rightarrow H)$ were estimated by simply taking the temperature difference between 0 and $H$ as shown in Fig. \ref{gd}(a). The results at 7.2 T [main panel of Fig. \ref{gd}(b)] are in reasonable agreement with the data estimated from the specific heat measurement (open squares)~\cite{dan2,kih}. Here, the data from Ref. 14 (open squares) shown in Fig. \ref{gd}(b) were interpolated from 7.5 T to 7.2 T. This agreement indicates the quantitative accuracy of our MCE measurements in the temperature range from 200 K to 320 K. In addition to this, we measured the $\Delta T_{\textrm{ad}}$ up to 55 T from $T_{\textrm{b}}$ data [Fig. \ref{gd}(c)] and extracted the temperature dependence of $\Delta T_{\textrm{ad}}(0\rightarrow20\textrm{T})$ and $\Delta T_{\textrm{ad}}(0\rightarrow55\textrm{T})$ as shown in the inset of Fig. \ref{gd}(b). We also obtained the similar temperature dependences of $\Delta T_{\textrm{ad}}$ with that of 7.2 T.

Figure \ref{gd}(c) displays the MCE up to 55 T at $T_{\textrm{b}}$ = 295 K. Here, the effect of magnetoresistance in the thermometer was compensated with using the data shown in Fig. \ref{au}. As discussed above, the artificial temperature change caused by the MR is less than 3 K in this temperature and field ranges, which are also shown in Fig. \ref{gd}(c) by the dotted curve. The observed $\Delta T_{\textrm{ad}}$ was substantially larger than the artificial effect and reaches 60 K at 55 T. Although such a huge temperature change might be felt unreasonable, we can prove its adequacy using the following simple calculation. With using Eq. \ref{dT} and Maxwell's relation $(\partial S(T,H)/\partial H)_T=(\partial M(T,H)/\partial T)_H$, we can calculate the MCE from the temperature dependence of magnetization $M$. Here we calculate the $M$ per mole using molecular filed approximation for ferromagnet:
\begin{align}
M &=Ng\mu_BJB_J\left(\frac{g\mu_BJ\mu_0\left(H-\lambda M\right)}{k_BT}\right), \\
\chi &=\frac{C}{T-\mathit{\Theta}}.
\end{align}
The $B_J$ is the Brillouin function. $\mu_0=4\pi \times 10^{-7}$ H/m is the space permeability. $\chi$ is the paramagnetic susceptibility. $C$ is the Curie constant [$C=Ng^2\mu_B^2J(J+1)/3k_B$]. $N\simeq6.02 \times 10^{23}$ mol$^{-1}$ is the Abogadro's number. Here, we set the parameters $g=2$ and $J=7/2$. The coefficient $\lambda$ determines the magnitude of the molecular field, and is estimated from the value of $T_{\textrm{C}}$ by $\lambda=\mathit{\Theta}/C\simeq T_{\textrm{C}}/C$. In the calculation of $\Delta T_{\textrm{ad}}$, we approximated the $C_H(T)$ as a constant value of 30 J/K. This value is the heat capacity of Gd at 350 K at zero field, which can be caused by lattice and electronic contribution~\cite{dan2}. The calculated result without any free parameter shows similar field dependence to the experimental result as shown in Fig. \ref{gd}(c). The $\Delta T_{\textrm{ad}}$ at 55 T is calculated to 62 K, which is 2 K larger than the experimental value. This difference of 2 K may be due to the underestimation of the heat capacity at zero field and/or the overestimation of the contribution of magnetoresistance of Au thermometer.

\begin{figure}[tbp]
\centering
\includegraphics[width=8.5cm,clip]{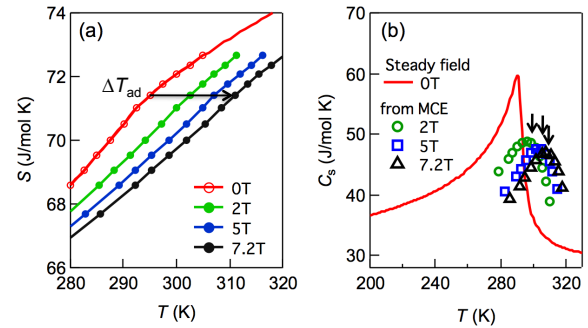}
\caption{(Color online) (a) Temperature dependence of entropy in polycrystalline Gd at 2 T, 5 T and 7.2 T determined by the isentropic shifts of entropies at zero field by $\Delta T_{\textrm{ad}}$. (b) Temperature dependence of the specific heats in Gd at 2 T, 5 T, and 7.2 T (open marks: estimated from MCEs, solid curves: measured in zero fields~\cite{dan2}).}
\label{c}
\end{figure}

Since the MCE is the phenomenon that reflects the change in the $S-T$ curve in zero- and finite-fields, we can, in principle, calculate the heat capacity at a certain field from the derivative of the $S-T$ curve in magnetic field determined using the MCE data. First, we calculate the $S-T$ curve with integrating the $C_H(T)$ measured at zero field. Since the total $S$ is conserved in the adiabatic condition, the entropy at a field $H$ is determined as $S(T+\Delta T_{\textrm{ad}},H)=S(T,0)$ by the experimental data of $\Delta T_{\textrm{ad}}$ as shown in Fig. \ref{c}(a). Finally, the $C_H(T)$ can be calculated from the derivative of this entropy curve. The results are shown in Fig. \ref{c}(b). The sharp peak at $T_{\textrm{C}}$ in zero field moves toward higher temperature by application of magnetic fields. Therefore, we can trace the field dependence of the second order transitions using this method.

To check the availability of our apparatus at the low temperature region, we measured the adiabatic MCE in GGG between 6 K and 30 K.
GGG is a garnet type compound with a cubic crystal structure, in which the magnetic Gd ions are on two corner-sharing triangular sublattices. Because of the strong geometric frustration in a garnet structure, the long-range ordering of the Gd spins ($J=7/2$) are suppressed to the extremely low temperature region ($<0.4$ K), and the paramagnetic property (Curie-Weiss behavior of magnetic susceptibility) are exhibited above 4.2 K~\cite{sch}. The MCE behavior in GGG have been studied in detail~\cite{lev}, and thus the GGG can be an ideal sample for testing the adiabatic MCE measurement system.

\begin{figure}[tbp]
\centering
\includegraphics[width=8.5cm,clip]{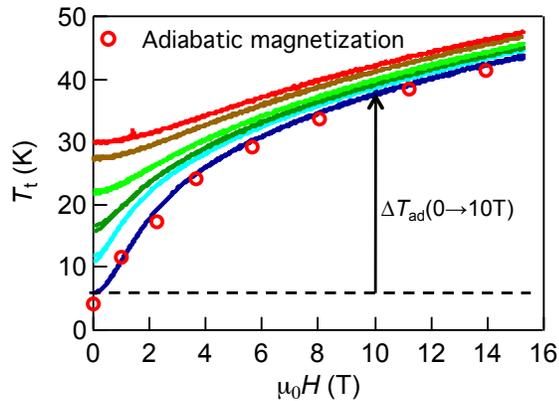}
\caption{(Color online) Field dependence of $T_{\textrm{t}}$ in Au$_{0.18}$Ge$_{0.82}$ film grown on the GGG crystal at the temperature of 6-30 K (solid line). Open circles: Field induced adiabatic temperature change of $T_{\textrm{s}}$\cite{lev}. The arrow corresponds to $\Delta T_{\textrm{ad}}(0\rightarrow10\textrm{T})$.}
\label{gg}
\end{figure}

Figure \ref{gg} shows the field dependence of the $T_{\textrm{t}}$. The observed reversible signals indicate that an adiabatic condition ($\tau_{\textrm{b}}\gg \tau_{\textrm{dur}}$) and a fast response of the thermometer ($\tau_{\textrm{dur}}\gg \tau_{\textrm{t}}$) are fulfilled. The positive MCE response is the typical behavior in a paramagnetic system, and the substantial size of the MCE is expected in Gd compounds having a large spin moment. For the quantitative comparison, the previous MCE data measured by an indirect MCE measurement are superimposed in Fig. \ref{gg} (open circles)~\cite{lev}, which shows a reasonable agreement with our data. This agreement confirms the validity of our MCE measurement technique down to 6 K. It should be noted that the indirect MCE measurement~\cite{lev} is only available when the calculation of the isothermal magnetization is valid. On the contrary, the direct measurement method of MCE is not necessary to estimate the isothermal magnetization curve and can be applied to any realistic system.

\begin{figure}
\centering
\includegraphics[width=8.5cm,clip]{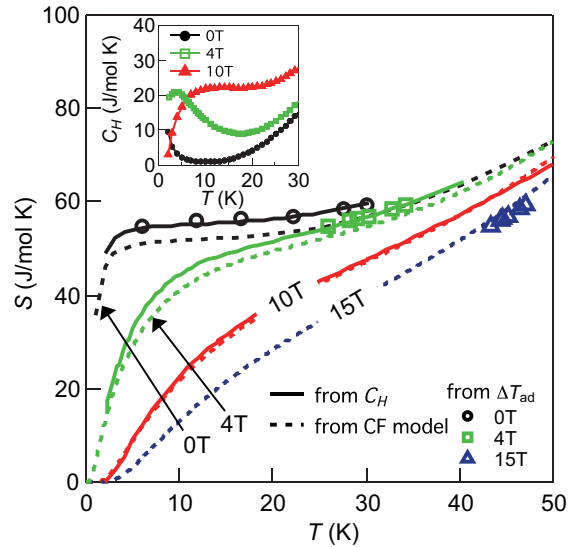}
\caption{(Color online) Entropy vs temperature in several magnetic fields. Solid curves: Entropies estimated from specific heats at 0 T, 4 T and 10 T, Dashed line: Entropies calculated from the crystal field model explained by Dai \textit{et al}~\cite{dai}, Open symbols: Entropies estimated from $\Delta T_{\textrm{ad}}$. Inset: $C_H-T$ curves at 0 T, 4 T and 10 T.}
\label{s}
\end{figure}

The inset of Fig. \ref{s} shows the temperature dependence of the specific heat measured on Quantum Design PPMS. The broad bump of $C_H$ observed below 20 K has been ascribed to the thermal excitation among the spin states~\cite{fis,dau,dai}. The application of magnetic fields enlarges the energy gaps among spin states, and the peak temperature of the bump moves from low to high temperature as the magnetic field increases. Since our $C_H$ data at 0 T and 4 T could not detect the large amount of entropy below 2 K, we estimated the $S-T$ curve at 10 T from the integration of the $C_H/T$ curve which is shown in Fig. \ref{s} (red solid curve). The curve is matched with the $S-T$ curve calculated by the crystal field model (dashed curve)~\cite{dai}. We take the $S-T$ curve at 10 T as a reference and estimate the temperatures at which the total entropy is the same as that of 10 T by using the data sets of $\Delta T_{\textrm{ad}}(10\textrm{T}\rightarrow H)$. The entropy data estimated by the MCE are plotted in Fig. \ref{s} as open circles (0 T), squares (4 T), and triangles (15 T). The low field $S-T$ curves estimated by the integration of $C_H/T$ are vertically shifted to be matched with the entropy estimated by the MCE and are also shown in Fig. \ref{s} by the solid curves. To check the validity of the method described above, we calculated $S-T$ curves by using a phenomenological model provided by Dai \textit{et al}~\cite{dai}. The calculated $S-T$ curve by this model are shown as dashed curves in Fig. \ref{s}. Our experimental results coincide with the calculated curves at 10 T and 15 T. On the other hand, there are tiny deviations from the experimental data at 0 T and 4 T. However, the calculation is also known to underestimate the zero-field entropy comparing to the experimental data obtained by the heat capacity measurement down to 75 mK~\cite{dun}. The prior study of heat capacity reported that the entropy at 2 K is $S=$ 47.7 J/mol K, which is almost the same as our estimated value of $S\simeq 49$ J/mol K at 2 K. These agreements confirm that the approach described above can be used to determine the absolute value of the entropy and its temperature dependence.

The specific heat measurement is known as a traditional method for the estimation of entropy. Although the entropy can be estimated throughout the measurement temperature range, it is difficult to obtain the absolute value of the entropy in materials that have large $C_H$ and/or residual entropy at the extremely low temperature. However these types of magnetic entropies is often released by applying appropriate magnetic fields, and the absolute value of the entropy can be obtained by tuning the strength of the magnetic field. The important point is that the adiabatic MCE can draw the isentropic curve as a function of $H$ and enables to deduce the absolute value of the entropy without the experiment at the extremely low temperature. In fact, reliable $S-T$ curves are obtained in Fig. \ref{s} without any heat capacity and MCE measurements below 2 K. Therefore, the adiabatic measurement of the MCE may open a unique opportunity to investigate the thermal properties of novel ground states.

\section{Conclusions}
We developed an adiabatic measurement system of the magneto-caloric effects (MCEs) in pulsed high magnetic fields up to 55 T. With preparing a thin film thermometer directly on the top of the sample surface, we achieved short response time sufficient for the measurement of the MCEs in pulsed fields. The obtained results on two samples of Gd and Gd$_3$Ga$_5$O$_{12}$ quantitatively agreed with the reported results. These results demonstrate the adequacy of our system in wide rages of temperatures and magnetic fields. We also estimated the heat capacity of the sample as a function of temperature by using the MCE results. The estimation can be useful to trace the change in critical temperatures of various phase transitions induced by magnetic fields.

\begin{acknowledgments}
This work was partly supported by the ministry of Education, Culture, Sports, Science and Technology, Japan, through a Grant-in-Aid for Scientific Research (23340096).
\end{acknowledgments}


\end{document}